\documentclass[11pt]{article}

\def\mytitle#1{\setcounter{equation}{0}
\setcounter{footnote}{0}
\begin{flushleft}\Large\textbf{#1}\end{flushleft}
\vspace{0.25cm}}
\def\myname#1{\leftline{{\large #1}}\vspace{-0.13cm}}
\def\myplace#1#2{\small\begin{flushleft}\textit{#1}\\
\texttt{#2}\end{flushleft}}

\usepackage{graphicx}
\begin{document}

\mytitle{Does \textit{f(R,T)} gravity admit a stationary scenario
between dark energy and dark matter in its framework?}

\vskip0.2cm \myname{Prabir Rudra\footnote{prudra.math@gmail.com}}
\myplace{Department of Mathematics, Indian Institute of
Engineering Science and Technology, Shibpur, Howrah-711 103, India.\\
Department of Mathematics, Asutosh College, Kolkata-700 026,
India.}{}

\begin{abstract}
In this note we address the well-known cosmic coincidence problem
in the framework of the \textit{f(R,T)} gravity. In order to
achieve this, an interaction between dark energy and dark matter
is considered. A constraint equation is obtained which filters the
\textit{f(R,T)} models that produce a stationary scenario between
dark energy and dark matter. Due to the absence of a universally
accepted interaction term introduced by a fundamental theory, the
study is conducted over three different forms of chosen
interaction terms. As an illustration three widely known models of
\textit{f(R,T)} gravity are taken into consideration and used in
the setup designed to study the problem. The study reveals that,
the realization of the coincidence scenario is almost impossible
for the popular models of $f(R,T)$ gravity, thus proving to be a
major setback for these models.
\end{abstract}

\vspace{5mm}


\vspace{5mm}

{\it Pacs. No.: 95.36.+x, 95.35.+d}\\

\vspace{5mm}

\section{Introduction}
Observational evidences from Ia supernovae, CMBR via WMAP, galaxy
redshift surveys via SDSS indicated that the universe have entered
a phase of accelerated expansion of late \cite{Perlmutter1,
Spergel1, Bennett1, Tegmark1, Allen1}. With this discovery the
incompleteness of general relativity (GR) as a self sufficient
theory of gravity came into foreground. Since no possible
explanation of this phenomenon could be attributed inside the
framework of Einstein's GR, a proper modification of the theory
was required that will successfully incorporate the late cosmic
acceleration. As the quest began, two different approaches
regarding this modification came into light.

According to the first approach, cosmic acceleration can be
phenomenally attributed to the presence of a mysterious negative
energy component popularly known as \textit{dark energy (DE)}
\cite{Riess1}. Here we modify the right hand side of the
Einstein's equation, i.e. in the matter sector of the universe.
Latest observational data shows that the contribution of DE to the
energy sector of the universe is $\Omega_{d}=0.7$. With the
passage of time, extensive search saw various candidates for DE
appear in the scene. Some of the popular ones worth mentioning are
Chaplygin gas models \cite{Kamenschik1, Gorini1}, Quintessence
Scalar field \cite{Ratra1}, Phantom energy field \cite{Caldwell1},
etc. A basic feature of these models is that, they violate the
strong energy condition i.e., $\rho+3p<0$, thus producing the
observed cosmic acceleration. Recent reviews on DE can be found in
\cite{Joyce1, Bam1}

A different section of cosmologists resorted to an alternative
approach for explaining the expansion. This concept is based on
the modification of the gravity sector of GR, thus giving birth to
modified gravity theories. A universe associated with a tiny
cosmological constant, i.e. the $\Lambda$CDM model served as a
prototype for this approach. It was seen that the model could
satisfactorily explain the recent cosmic acceleration and passed a
few solar system tests as well. But with detailed diagnosis it was
revealed that the model was paralyzed with a few cosmological
problems. Out of these, two major problems that crippled the model
till date are the Fine tuning problem (FTP) and the Cosmic
Coincidence problem (CCP). The FTP refers to the large discrepancy
between the observed values and the theoretically predicted values
of cosmological parameters. Numerous attempts to solve this
problem can be found in the literature. Among them, the most
impressive attempt was undertaken by Weinberg in \cite{Weinberg1}.
Although the approaches for the solutions are different, yet,
almost all of them are basically based on the fact that the
cosmological constant may not assume an extremely small static
value at all times during the evolution of the universe (as
predicted by GR), but its nature should be rather dynamical
\cite{Bisabr1}. These drawbacks reduced the effectiveness of the
model, as well as its acceptability, and hence alternative
modifications of gravity was sought for. Some of the popular
models of modified gravity that came into existence in recent
times are loop quantum gravity \cite{Rovelli1, Ashtekar1}, Brane
gravity \cite{Brax1, Maartens1, Maartens2}, \textit{f(R)} gravity
\cite{Kerner1, Allemandi1, Carroll1}, \textit{f(T)} gravity
\cite{Linder1, Li1, Miao1, Li2}, etc. Reviews on extended gravity
theories can be found in \cite{Capo1, Noji1, Clifton1}.

In this work we will consider \textit{f(R,T)} model as the theory
of gravity \cite{Harko1}. Over the years, several modifications to
GR have been achieved, by generalizing the Einstein-Hilbert
Lagrangian used in GR. $f(R)$ and $f(T)$ gravities are common
examples of such modifications. In $f(R)$ gravity the Ricci scalar
$R$ is replaced by a general function of $R$ in the
Einstein-Hilbert action. Levi-Civita connection is used in the
theory with only curvature for its formation. In $f(T)$ gravity,
the torsion scalar $T$ is replaced by a general function of $T$ in
the action of the teleparallel equivalent of general relativity
(TEGR). Weitzenbock connection is used in the theory where the
curvature is replaced by torsion. Both these theories can
successfully explain the recent cosmic acceleration and passes
several solar system tests \cite{Bisabr, Jun, Berry, Xie, Iorio}.

$f(R,T)$ gravity is a novel attempt to unify the $f(R)$ and $f(T)$
gravities preserving the properties of the constituent theories.
The evolution of the universe is explained under the combined
effect of both curvature scalar, $R$ and torsion, $T$. Over the
past two years $f(R,T)$ gravity has evolved as a prospective and a
very interesting version of modified gravity theory. Since the
introduction of the theory, numerous works have been recorded in
the literature, investigating its various aspects. The
thermodynamic properties of the theory are studied in
\cite{Sharif1}. The energy conditions are studied in
\cite{Alvarenga1}. Anisotropic cosmology in the background of
\textit{f(R,T)} gravity was studied in \cite{Sharif2}.
Cosmological solution via reconstruction program was studied in
\cite{Jamil1}.

In \cite{Bisabr2} Bisabr studied cosmological coincidence problem
in the background of \textit{f(R)} gravity. In \cite{Rudra2} and
\cite{Rudra3}, the problem has been studied in $f(G)$ and $f(T)$
gravities respectively. Motivated by these, we dedicate the
present assignment to the study of the coincidence problem in
\textit{f(R,T)} gravity. The paper is organized as follows: Basic
equations of \textit{f(R,T)} gravity are furnished in section 2.
In section 3, we discuss the coincidence problem. The set-up for
the present study is discussed in section 4. We illustrate the
designed set-up by a few examples in section 5, and finally the
paper ends with a short conclusion in section 6.

\section{Basic Equations of \textit{f(R,T)} Gravity}
The action of the \textit{f(R,T)} gravity theory is given by
\begin{equation}
A=\int dx^{4}\sqrt{-g}\left[\frac{f(R,T)}{16\pi G}+\cal{L}\right]
\end{equation}
where $\cal{L}$ determines the matter content of the universe. The
energy momentum tensor of matter is defined as
\begin{equation}
T_{\mu\nu}^{(matter)}=-\frac{2}{\sqrt{-g}}\frac{\delta(\sqrt{-g}\cal{L})}{\delta
g^{\mu\nu}}
\end{equation}
We assume that the matter Lagrangian density depends only on the
metric tensor components $g_{\mu\nu}$ so that
\begin{equation}
T_{\mu\nu}^{(matter)}=g_{\mu\nu}L_{(matter)}-\frac{2\delta
L_{(matter)}}{\delta g^{\mu\nu}}
\end{equation}
Taking the variation of the action with respect to the metric
tensor, we get the field equations for the \textit{f(R,T)} gravity
as follows,
\begin{equation}
R_{\mu\nu}f_{R}(R,T)-\frac{1}{2}g_{\mu\nu}f(R,T)+\left(g_{\mu\nu}~{\fbox{}}-\nabla_{\mu}\nabla_{\nu}\right)f_{R}(R,T)=8\pi
G
T_{\mu\nu}^{(matter)}-f_{T}(R,T)T_{\mu\nu}^{(matter)}-f_{T}(R,T)\Theta_{\mu\nu}
\end{equation}
where $\nabla_{\mu}$ is the covariant derivative associated with
the Levi-Civita connection of the metric and
$\fbox{}=\nabla_{\mu}\nabla^{\mu}$. Subscripts $R$ and $T$
represents derivative with respect to $R$ and $T$ respectively and
$\Theta_{\mu\nu}=\frac{g^{\alpha\beta}\delta
T_{\alpha\beta}}{\delta g^{\mu\nu}}$.

The energy-momentum tensor of the matter is given as
\begin{equation}
T_{\mu\nu}^{(matter)}=\left(\rho_{m}+p_{m}\right)u_{\mu}u_{\nu}+p_{m}g_{\mu\nu}
\end{equation}
Using this the field eqns. (4) gives,
\begin{equation}
R_{\mu\nu}f_{R}(R,T)-\frac{1}{2}g_{\mu\nu}f(R,T)+\left(g_{\mu\nu}\fbox{}\nabla_{\mu}\nabla_{\nu}\right)f_{R}(R,T)=8\pi
G
T_{\mu\nu}^{(matter)}+T_{\mu\nu}^{(matter)}f_{T}(R,T)+p_{m}g_{\mu\nu}f_{T}(R,T)
\end{equation}
Here we will consider non-relativistic matter, i.e., cold dark
matter and baryons $(p_{m}=0)$. So its obvious that the torsion
contribution is totally coming from ordinary matter. An effective
Einstein eqn. can be written from eqn.(6) as follows,
\begin{equation}
R_{\mu\nu}-\frac{1}{2}R g_{\mu\nu}=8\pi
G_{eff}T_{\mu\nu}^{(matter)}+T_{\mu\nu}^{(d)}
\end{equation}
where the effective gravitational matter dependant coupling in
$f(R,T)$ gravity is given by,
\begin{equation}
G_{eff}=\frac{1}{f_{R}(R,T)}\left(G+\frac{f_{T}(R,T)}{8\pi}\right)
\end{equation}
and the energy momentum tensor for dark energy is given by,
\begin{equation}
T_{\mu\nu}^{(d)}=\frac{1}{f_{R}(R,T)}\left[\frac{1}{2}g_{\mu\nu}\left(f(R,T)-Rf_{R}(R,T)\right)+
\left(\nabla_{\mu}\nabla_{\nu}-g_{\mu\nu}\fbox{}\right)f_{R}(R,T)\right]
\end{equation}
here prime denotes the non-equilibrium description of the field
equations.

Now the metric describing the FRW universe is given by,
\begin{equation}
ds^{2}=h_{\alpha\beta}dx^{\alpha}dx^{\beta}+\tilde{r}^{2}d\Omega^{2}
\end{equation}
where $\tilde{r}=a(t)r$, $x^{0}=t$, $x^{1}=r$ and the metric
$h_{\alpha\beta}=diag(-1,\frac{a^{2}}{1-kr^{2}})$. Obviously
$a(t)$ is the time dependant scale factor, $k$ is the scalar
curvature and $d\Omega^{2}$ is the metric for the two-dimensional
unit sphere. Now the field equations for the FRW universe can be
given as follows,
\begin{equation}
3\left(H^{2}+\frac{k}{a^{2}}\right)=8\pi
G_{eff}\rho_{m}+\frac{1}{f_{R}}\left[\frac{1}{2}\left(Rf_{R}-f\right)-3H\left(\dot{R}f_{RR}+\dot{T}f_{RT}\right)\right]
\end{equation}
\begin{equation}
-\left(2\dot{H}+3H^{2}+\frac{k}{a^{2}}\right)=\frac{1}{f_{R}}\left[-\frac{1}{2}\left(Rf_{R}-f\right)
+2H\left(\dot{R}f_{RR}+\dot{T}f_{RT}\right)+\ddot{R}f_{RR}+\dot{R}^{2}f_{RRR}+2\dot{R}\dot{T}f_{RRT}
+\ddot{T}f_{RT}+\dot{T}^{2}f_{RTT}\right]
\end{equation}
We re-write these eqns. as
\begin{equation}
3\left(H^{2}+\frac{k}{a^{2}}\right)=8\pi
G_{eff}\left(\rho_{m}+\rho_{d}\right)
\end{equation}
\begin{equation}
-2\left(\dot{H}-\frac{k}{a^{2}}\right)=8\pi
G_{eff}\left(\rho_{m}+\rho_{d}+p_{d}\right)
\end{equation}
where $\rho_{d}$ and $p_{d}$ are respectively the energy density
and pressure of dark energy given by,
\begin{equation}
\rho_{d}=\frac{1}{8\pi G
F}\left[\frac{1}{2}\left(Rf_{R}-f\right)-3H\left(\dot{R}f_{RR}+\dot{T}f_{RT}\right)\right]
\end{equation}
\begin{equation}
p_{d}=\frac{1}{8\pi G F}\left[-\frac{1}{2}\left(Rf_{R}-f\right)
+2H\left(\dot{R}f_{RR}+\dot{T}f_{RT}\right)+\ddot{R}f_{RR}+\dot{R}^{2}f_{RRR}+2\dot{R}\dot{T}f_{RRT}
+\ddot{T}f_{RT}+\dot{T}^{2}f_{RTT}\right]
\end{equation}
Here $F=1+\frac{f_{T}(R,T)}{8\pi G}$.

The energy conservation equations for matter and dark sector are
respectively given by,
\begin{equation}
\dot{\rho_{m}}+3H\rho_{m}=Q
\end{equation}
and
\begin{equation}
\dot{\rho_{d}}+3H\left(1+\omega_{d}\right)\rho_{d}=-Q
\end{equation}

Here $\omega_{d}=\frac{p_{d}}{\rho_{d}}$ is the EoS parameter of
the energy sector and $Q$ is the interaction between the matter
and the energy sector of the universe. The EoS parameter of the
dark fluid is obtained as,
\begin{equation}
\omega_{d}=-1+\frac{\ddot{R}f_{RR}+\dot{R}^{2}f_{RRR}+2\dot{R}\dot{T}f_{RRT}
+\ddot{T}f_{RT}+\dot{T}^{2}f_{RTT}-H\left(\dot{R}f_{RR}+\dot{T}f_{R,T}\right)}{\frac{1}{2}\left(Rf_{R}
-f\right)-3H\left(\dot{R}f_{RR}+\dot{T}f_{RT}\right)}
\end{equation}

\section{The Coincidence problem}

The cosmic coincidence problem has been a serious issue in recent
times regarding various otherwise successful models of the
universe. From the recent cosmological observations it is noted
that the densities of the matter sector and the DE sector of the
universe are almost identical in late times. This observation
gives rise to a problem when we relate this to the fact that the
matter and the energy component of the universe have evolved
independently from different mass scales in the early universe.
Then how do they reconcile to identical mass scales in the late
universe! This is a major cosmological problem having its roots in
the very formation of the models. Almost all the models of
universe known till date more or less suffer from this phenomenon.

Numerous attempts to address the coincidence problem can be widely
found in literature. Among them the most impressive are the ones
which introduce a suitable interaction between the matter and the
energy components of the universe, as given in the conservation
equations (17) and (18). This approach makes use of the fact that
the two sectors of the universe have not evolved independently
from different mass scales, but have actually evolved together,
interacting with each other, thus allowing a mutual flow of matter
and energy between the two components. Due to this exchange,
difference of densities, if any, gets diluted and a stationary
scenario is witnessed in the present universe. Although the
concept seems to be a really promising one, yet a problem
persists. Till date there is no universally accepted interaction
term, introduced by a fundamental theory. An attempt to address
the coincidence problem in $f(R)$ gravity can be found in
\cite{Bisabr2}. Similar attempts in $f(G)$ and $f(T)$ gravities
can be found in \cite{Rudra2, Rudra3} respectively. A study of
triple interacting DE model can be found in \cite{Huang1}.

It is known that both dark energy and dark matter are not
universally accepted facts, but concepts which are still at the
speculation level. Due to this unknown nature of both dark energy
and dark matter, it is not possible to derive an expression for
the interaction term ($Q$) from the first principles. Such a
situation, demands us to use our logical reasoning and propose
various expressions for $Q$ that will be reasonably acceptable.
The late time dominating nature of dark energy indicates that $Q$
must be considered a small and positive value. On the other hand a
large negative value of interaction will make the universe dark
energy dominated from the early times, thus leaving no scope for
the condensation of galaxies. So the most logical choice for
interaction should contain a product of energy density and the
hubble parameter, because it is not only physically but also
dimensionally justified. So $Q=Q(H\rho_{m}, H\rho_{de})$, where
$\rho_{de}$ is the dark energy density. Since here we are not
planning to add any dark energy by hand, so the effective density
resulting from the \textit{f(R,T)} gravity, $\rho_{d}$ will
replace $\rho_{de}$. This leads us to three basic forms of
interactions as given below \cite{del Campo1}:
\begin{equation}
b-model:~ Q=3b H\rho_{m}~~~~~~~~~\eta-model:~ Q=3\eta
H\rho_{d}~~~~~~~~\Gamma-model:~ Q=3\Gamma
H\left(\rho_{m}+\rho_{d}\right),~~~~~~~~
\end{equation}
where $b$, $\eta$ and $\Gamma$ are the coupling parameters of the
respective interaction models.

It is worth mentioning that due to its simplicity as well as
viability, the most widely used interaction model is the $b$-model
and is available widely in literature \cite{Berger1, del Campo1,
Rudra1, Jamil2}.

\section{The set-up}
In this note we address the coincidence problem in \textit{f(R,T)}
gravity. \textit{f(R,T)} gravity has evolved over the past few
years as a candidate for modified gravity theory. From the
literature it is known that \textit{f(R,T)} gravity is itself self
competent in producing the late cosmic acceleration without
resorting to any forms of dark energy. Therefore in order to keep
it simple and reasonable, we do not consider any separate dark
energy component by hand in the present study. The energy
component evolving from the gravity theory itself is considered as
the dark component responsible for the cosmic acceleration. The
ratio of the densities of matter and dark energy is considered as,
$r\equiv \rho_{m}/\rho_{d}$. Our prime objective is to devise a
set-up that will take us close to a possible solution of the
coincidence problem. We also want to set up a filtering process
that will screen the favorable $f(R,T)$ models, that produce a
stationary scenario of the component densities, $r$ from the
unfavorable ones which do not. The time evolution of $r$ is as
follows,
\begin{equation}
\dot{r}=\frac{\dot{\rho_{m}}}{\rho_{d}}-r\frac{\dot{\rho_{d}}}{\rho_{d}}
\end{equation}
Using eqns. (17), (18) and (21), we obtain
\begin{equation}
\dot{r}=3Hr\omega_{d}+\frac{Q}{\rho_{d}}\left(1+r\right)
\end{equation}

Using the b-interaction given in eqn.(20), we get the expression
for $\dot{r}$ as,
\begin{equation}
\dot{r}=3Hr\left(b+br+\omega_{d}\right)
\end{equation}
where $\omega_{d}$ is given by eqn.(19). Now in order to comply
with observations, it is required that universe should approach a
stationary stage, where either $r$ becomes a constant or evolves
slower than the scale factor. In order to satisfy this $\dot{r}=0$
in the present epoch, it leads to the following equation,
\begin{equation}
g_{1}(f,H,r_{s},q)=0
\end{equation}
where

$$g_{1}(f,H,r_{s},q)=3Hr_{s}
\left[b+\frac{{\dot{H}}}{H^2}+q+br_{s}+\frac{1}{\frac{1}{2}\left(-f+Rf_{R}-3H\left(6\left(\ddot{H}+4H\dot{H}\right)
f_{RR}-12H\dot{H}f_{RT}\right)\right)}\times\right.$$

$$\left.\left(6\left(\stackrel{...}H+4\left(\dot{H}^{2}+H\ddot{H}\right)\right)f_{RR}-12\left(\dot{H}^{2}+H\ddot{H}\right)f_{RT}
-H\left(6\left(\ddot{H}+4H\dot{H}\right)f_{RR}-12H\dot{H}f_{RT}+
36\left(\ddot{H}+4H\dot{H}\right)^{2}f_{RRR}\right.\right.\right.$$

\begin{equation}
\left.\left.\left.-144H\dot{H}\left(\ddot{H}+4H\dot{H}\right)f_{RRT}+144\dot{H}^{2}H^{2}f_{RTT}\right)\right)\right]
\end{equation}
and $r_{s}$ is the value of $r$ when it takes a stationary value.

Using the $\eta$-interaction given in eqn.(20), we get the
expression for $\dot{r}$ as,
\begin{equation}
\dot{r}=3H\left[\eta+r\left(\eta+\omega_{d}\right)\right]
\end{equation}
where $\omega_{d}$ is given by eqn.(19). In order to satisfy this
$\dot{r}=0$ in the present epoch, it leads to the following
equation,
\begin{equation}
g_{2}(f,H,r_{s},q)=0
\end{equation}
where

$$g_{2}(f,H,r_{s},q)=3H \left[\eta+r_{s}
\left(\frac{\dot{H}}{H^2}+q+\eta+\frac{1}{\left(\frac{1}{2}\left(-f+6\left(\dot{H}+2H^{2}\right)f_{R}
-3H\left(6\left(\ddot{H}+4H\dot{H}\right)f_{RR}-12H\dot{H}f_{RT}\right)\right)\right)}\times\right.\right.$$

$$\left.\left.\left(6\left(\stackrel{...}H+4\left(\dot{H}^{2}+H\ddot{H}\right)\right)f_{RR}-12\left(\dot{H}^{2}
+H\ddot{H}\right)f_{RT}-H\left(6\left(\ddot{H}+4H\dot{H}\right)f_{RR}-12H\dot{H}f_{RT}\right)
+36\left(\ddot{H}+4H\dot{H}\right)^{2}f_{RRR}\right.\right.\right.$$

\begin{equation}
\left.\left.\left.-144H\dot{H}\left(\ddot{H}+4H\dot{H}\right)f_{RRT}+144\dot{H}^{2}H^{2}f_{RRT}\right)\right)\right]
\end{equation}

Using the $\Gamma$-interaction given in eqn.(20), we get the
expression for $\dot{r}$ as,
\begin{equation}
\dot{r}=3H\left[\Gamma
r^{2}+r\left(2\Gamma+\omega_{d}\right)+\Gamma\right]
\end{equation}
where $\omega_{d}$ is given by eqn.(19). In this case, in order to
satisfy $\dot{r}=0$ in the present epoch, it leads to the
following equation,
\begin{equation}
g_{3}(f,H,r_{s},q)=0
\end{equation}
where

$$g_{3}(f,H,r_{s},q)= 3H\left[\Gamma+r_{s}^2\Gamma+r_{s}
\left(\frac{\dot{H}}{H^2}+q+2\Gamma+
\frac{1}{\left(\frac{1}{2}\left(-f+Rf_{R}\right)-3H\left(6\left(\ddot{H}+4H\dot{H}\right)f_{RR}
-12\dot{H}Hf_{RT}\right)\right)}\times\right.\right.$$

$$\left.\left.\left(6\left(\stackrel{...}+4\left(\dot{H}^{2}+H\ddot{H}\right)\right)f_{RR}
-12\left(\dot{H}^{2}+H\ddot{H}\right)f_{RT}-H\left(6\left(\ddot{H}+4H\dot{H}\right)f_{RR}-12H\dot{H}f_{RT}+36
\left(\ddot{H}+4H\dot{H}\right)^{2}f_{RRR}\right.\right.\right.\right.$$

\begin{equation}
\left.\left.\left.\left.-144H\dot{H}\left(\ddot{H}+4H\dot{H}\right)f_{RRT}
+144\dot{H}^{2}H^{2}f_{RTT}\right)\right)\right)\right]
\end{equation}

In our analysis we will consider $H_{0}$, $r_{0}$ and $q_{0}$ as
the present day values of $H$, $r$ and $q$ respectively. As far as
$q$ is concerned, we start from the best fit parametrization
obtained directly from observational data. Here we use a two
parameter reconstruction function for $q(z)$ \cite{Gong1, Gong2}
\begin{equation}
q(z)=\frac{1}{2}+\frac{q_{1}z+q_{2}}{\left(1+z\right)^{2}}
\end{equation}
On fitting this model to Gold data set, we get
$q_{1}=1.47_{-1.82}^{+1.89}$ and $q_{2}=-1.46\pm 0.43$
\cite{Gong2}. We consider $z_{0}=0.25$ and using these values in
eqn.(32), we get $q_{0}\approx -0.2$. From recent observations, we
obtain $r_{0}\equiv \frac{\rho_{m}(z_{0})}{\rho_{T}(z_{0})}\approx
\frac{3}{7}$ \cite{Zlatev1, Wei1, Yang1}. The present value of
Hubble parameter, $H_{0}$ is taken as 72, in accordance with the
latest observational data.

\section{Illustration}

Here we consider the scale-factor, of the $\Lambda$CDM universe,
which describes the late universe quite satisfactorily. For a DE
component with EoS $w$, we have $\rho(t)\sim a^{-3(1+w)}$ and thus
$a(t)\sim t^{2/(3(1+w))}$. So in the early stage, where there is
only radiation $(w=1/3)$ then $a(t)\sim t^{1/2}$. In the late
universe, when we have only matter $a(t)\sim t^{2/3}$. Therefore
we consider our scale factor as,

\begin{equation}
a=a_{0}t^{2/3}
\end{equation}
where $a_{0}$ is a constant.

In order to illustrate the above set-up numerically we consider
three different $f(R,T)$ gravity models found in literature and
test them for the coincidence phenomenon. These models are used
because they pass most of the cosmological and solar system tests.
The three models are \cite{Shabani}\\

\vspace{5mm} ~~~~~~~~~~~~~~\textbf{Model1:}
\begin{equation}\label{24}
f(R,T)=\mu R^{n_{1}}+\nu T^{m}
\end{equation}
where $\mu$, $n_{1}$, $\nu$ and $m$ are constants.

\vspace{5mm}~~~~~~~~~~~~~~\textbf{Model2:}
\begin{equation}\label{25}
f(R,T)=R^{p}\left(Log(\alpha R)\right)^{q_{3}}+\sqrt{-T}
\end{equation}
where $p$, $\alpha>0$ and $q_{3}\neq 0$  are constants.

\vspace{5mm}~~~~~~~~~~~~~~\textbf{Model3:}
\begin{equation}\label{26}
f(R,T)=R+\beta R^{-n_{2}}+\sqrt{-T}
\end{equation}
where $n_{2}\neq 0$ and $\beta$ are constants.

Using the model 1, i.e., eqn.(34) and eqn. (33) in eqn. (25), we
get the following expression for the dynamical quantity $g_{1}$,

$$g_{1}^{model1}=\frac{1}{t}\left[3nr\left(b-\frac{1}{n}+q+br+\left\{-\frac{2^{-1+n_{1}}
3^{n_{1}} n (-1+n_{1})n_{1} \left(\frac{2n}{t^{3}}-\frac{4
n^{2}}{t^{3}}\right) \left(-\frac{n}{t^2}+\frac{2
n^2}{t^2}\right)^{-2+n_{1}}\mu}{t}\right.\right.\right.$$

$$\left.\left.\left.+\frac{1}{2} \left(-6^{n_{1}} \left(-\frac{n}{t^2}+\frac{2
n^2}{t^2}\right)^{n_{1}}\mu +6^{n_{1}} n_{1}
\left(-\frac{n}{t^2}+\frac{2 n^2}{t^2}\right)^{n_{1}} \mu -6^{m}
\left(-\frac{n^2}{t^2}\right)^{m} \nu
\right)\right\}^{-1}\right.\right.\times$$

$$\left.\left.\left(6^{-1+n_{1}} \left(-2+n_{1}\right)
\left(-1+n_{1}\right) n_{1} \left(\frac{2 n}{t^3}-\frac{4
n^2}{t^3}\right)^2 \left(-\frac{n}{t^2}+\frac{2
n^2}{t^2}\right)^{-3+n_{1}}
\mu+6^{-1+n_{1}}\left(-1+n_{1}\right)n_{1}\right.\right.\right.$$

\begin{equation}
\left.\left.\left. \left(-\frac{6 n}{t^4}+\frac{12
n^2}{t^4}\right) \left(-\frac{n}{t^2}+\frac{2
n^2}{t^2}\right)^{-2+n_{1}} \mu-\frac{6^{-1+n_{1}} n
\left(-1+n_{1}\right) n_{1} \left(\frac{2 n}{t^3}-\frac{4
n^2}{t^3}\right) \left(-\frac{n}{t^2}+\frac{2
n^2}{t^2}\right)^{-2+n_{1}}\mu}{t}\right)\right)\right]
\end{equation}

Similarly expressions for $g_{1}$ is obtained for the other two
$f(R,T)$ models. Expressions for $g_{2}$ and $g_{3}$ are also
found for all the three gravity models. As it can be seen from
above that the expressions are really lengthy, so we do not
include all of them in the manuscript.

We have generated plots for $g_{1}$, $g_{2}$ and $g_{3}$ against
cosmic time, $t$ for each of the three models in the figures 1, 2
and 3, for all the three forms of interactions, $b$, $\eta$ and
$\Gamma$ respectively. Particular numerical values for the
involved parameters have been considered which are in accordance
with the recent observational data \cite{Nojiri1, Bamba2,
Schmidt1, Shabani, Sharif3}.

\vspace{2mm}
\begin{figure}
~~~~~~~~~~~~~~~~~\includegraphics[height=3in]{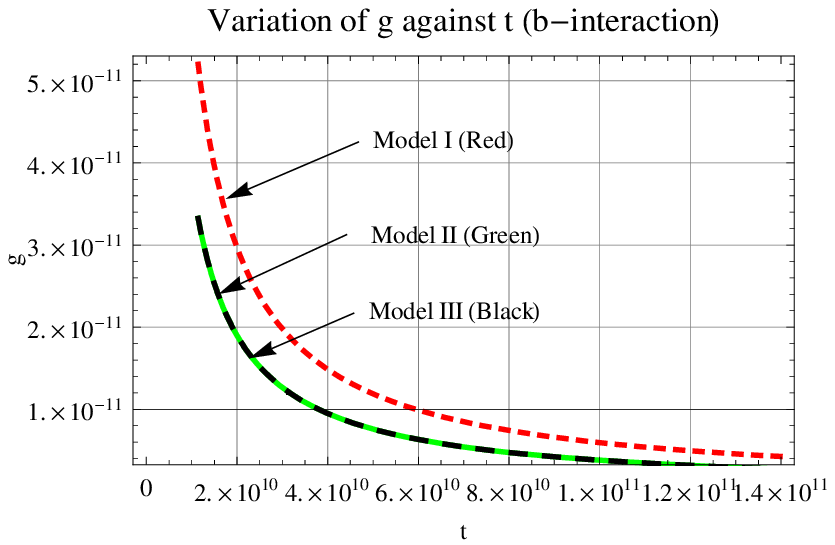}~~~~\\
\vspace{5mm}
~~~~~~~~~~~~~~~~~~~~~~~~~~~~~~~~~~~~~~~~~~~~~~~~~~~~~~~Fig.1~~~~~~~~~~~~~~~~~~~~~~~~~~~~~~~~~\\
\vspace{1mm} \textsl{Fig 1 : The plot of
$g(f_{0},H_{0},r_{s0},q_{0})$ against $t$ for model1 (red), model2
(green) and model3 (black) using $b$ interaction. The other
parameters are considered as $q=-0.2, r=3/7, b=1.5, \mu=0.01,
\nu=0.05, n_{1}=0.5, m=3, \alpha=5.7\times 10^{-61}, p=1,
q_{3}=-1,
n_{2}=-0.9, \beta=0.646\times 10^{-4}$}\\
\end{figure}

\vspace{3mm}

\begin{figure}
~~~~~~~~~~~~~~~~~\includegraphics[height=3in]{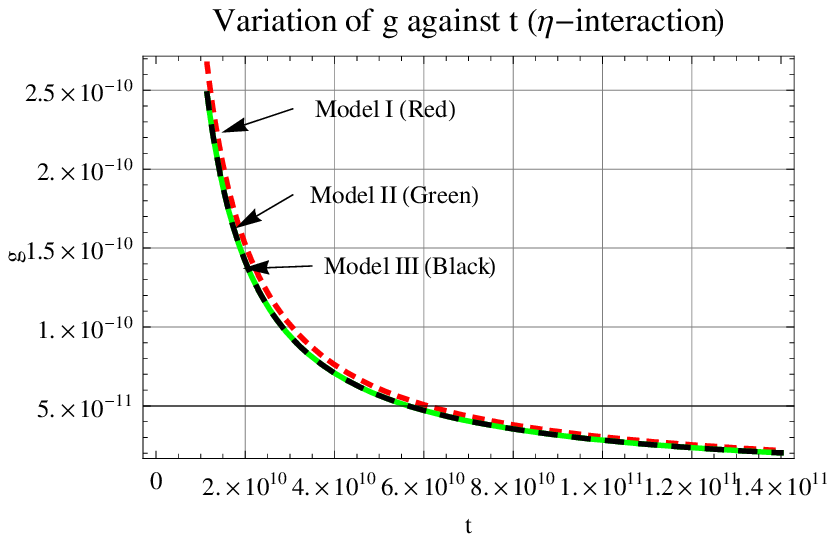}~~~~\\
\vspace{5mm}
~~~~~~~~~~~~~~~~~~~~~~~~~~~~~~~~~~~~~~~~~~~~~~~~~~~~~~~Fig.2~~~~~~~~~~~~~~~~~~~~~~~~~~~~~~~~~\\
\vspace{1mm} \textsl{Fig 2 : The plot of
$g(f_{0},H_{0},r_{s0},q_{0})$ against $t$ for model1 (red), model2
(green) and model3 (black) using $\eta$ interaction. The other
parameters are considered as $q=-0.2, r=3/7, b=1.5, \mu=0.01,
\nu=0.05, n_{1}=0.5, m=3, \alpha=5.7\times 10^{-61}, p=1,
q_{3}=-1,
n_{2}=-0.9, \beta=0.646\times 10^{-4}$}\\
\end{figure}

\vspace{2mm}
\begin{figure}
~~~~~~~~~~~~~~~~~\includegraphics[height=3in]{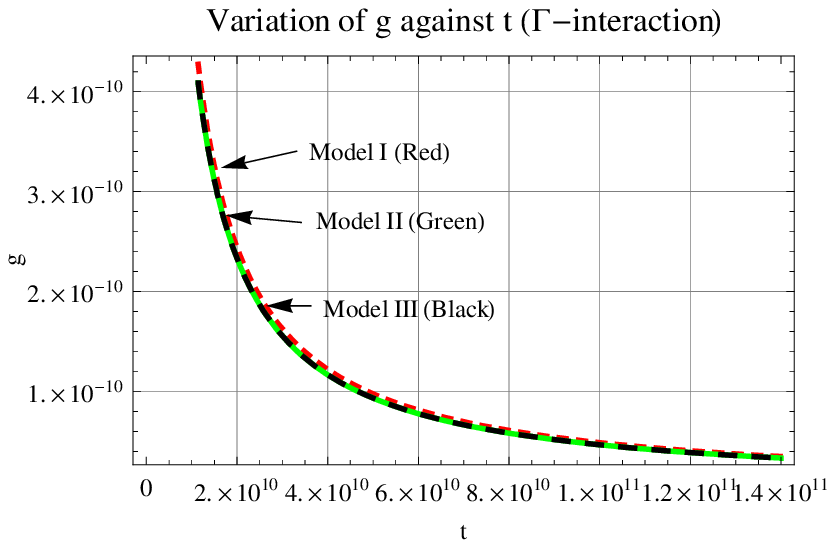}~~~~\\
\vspace{5mm}
~~~~~~~~~~~~~~~~~~~~~~~~~~~~~~~~~~~~~~~~~~~~~~~~~~~~~~~Fig.3~~~~~~~~~~~~~~~~~~~~~~~~~~~~~~~~~\\
\vspace{1mm} \textsl{Fig 3 : The plot of
$g(f_{0},H_{0},r_{s0},q_{0})$ against $t$ for model1 (red), model2
(green) and model3 (black) using $\Gamma$ interaction. The other
parameters are considered as $q=-0.2, r=3/7, b=1.5, \mu=0.01,
\nu=0.05, n_{1}=0.5, m=3, \alpha=5.7\times 10^{-61}, p=1,
q_{3}=-1,
n_{2}=-0.9, \beta=0.646\times 10^{-4}$}\\
\end{figure}

\section{Discussion and Conclusion}
From the figures it is evident that the stationary scenario is not
realized for all the three models when the cosmic time corresponds
to the age of the universe, i.e. $14\times 10^{9}$ years. It is
seen that near the zero line the trajectories assume asymptotic
nature. Thus there is no realistic possibility of the trajectory
to intersect the zero axis, producing a non-stationary scenario.
The asymptotic nature of the curves are indicative of the fact
that as time evolves the trajectories move closer and closer to
the zero mark thus alleviating the coincidence problem
substantially, but never produces a satisfactory solution. This
reveals a basic flaw in the framework of the $f(R,T)$ models which
are otherwise considered to be quite consistent with the solar
system tests. In the figures 1, 2 and 3 the trajectories have been
generated for $b$, $\eta$ and $\Gamma$-interaction respectively.
In all the three cases, the trajectories for model 2 and 3
coincide with each other. The trajectory for model 1 is distinct
from the other models in case of b-interaction. But in case of
$\eta$ and $\Gamma$ interactions even the trajectory for model 1
move closer to the other trajectories almost coinciding with them.

In \cite{Rudra2, Rudra3} coincidence problem has been addressed in
$f(G)$ and $f(T)$ gravity models respectively. In both these
assignments the b-interaction has been identified as the most
suitable form of interaction describing the late universe. But in
the present study, there is no such reason to consider the above
mentioned fact. But it should be mentioned that the present study
resembles the study in \cite{Rudra2} in the fact that in both the
papers coincidence scenario is not realized for the well-known
models of the respective gravity theories. Finally it must be
stated that from the set-up that we have designed in this
assignment, we can generate as well as filter various models of
$f(R,T)$ gravity which are completely free from the coincidence problem.
For the time being we keep it as a future assignment.\\

\section*{Acknowledgements}

The author sincerely thanks E.N. Saridakis for helpful
discussions. The Author also acknowledges the anonymous referee
for enlightening comments that helped to improve the quality of the manuscript.\\

\end{document}